\documentclass[journal=jacsat,manuscript=article,layout=twocolumn]{achemso}
\usepackage{amssymb,amsmath,amsthm,amsfonts,xcolor}
\usepackage{tablefootnote}
\usepackage{multicol}
\usepackage[mathlines]{lineno}

\usepackage{xr}
\makeatletter
    \newcommand*{\addFileDependency}[1]{
    \typeout{(#1)}
    \@addtofilelist{#1}
    \IfFileExists{#1}{}{\typeout{No file #1.}}
    }
\makeatother
\newcommand*{\myexternaldocument}[1]{
    \externaldocument{#1}
    \addFileDependency{#1.tex}
    \addFileDependency{#1.aux}
}
\myexternaldocument{SI_methods_JS_flag}
\myexternaldocument{SI_protocol_JS_flag}
\myexternaldocument{SI_data_JS_flag}

\graphicspath{{Figures_flag/}}
\newcommand{\IC}{IC$_\text{50}$} 
\newcommand{\Ki}{$K_i$}

\title{Drug Resistance Predictions Based on a Directed Flag Transformer}

\author{Dong Chen} 
\affiliation{
Department of Mathematics, Michigan State University, MI, 48824, USA
}
\author{Gengzhuo Liu} 
\affiliation{
Department of Mathematics, Michigan State University, MI, 48824, USA
}
\author{Hongyan Du} 
\affiliation{
Department of Mathematics, Michigan State University, MI, 48824, USA
}
\author{Benjamin Jones} 
\affiliation{
Department of Mathematics, Michigan State University, MI, 48824, USA
}

\author{Junjie Wee} 
\affiliation{
Department of Mathematics, Michigan State University, MI, 48824, USA
}
\author{Rui Wang} 
\affiliation{
Simons Center for Computational Physical Chemistry, New York University, New York, NY, 10003
}
\author{Jiahui Chen} 
\affiliation{
Department of Mathematical Sciences, University of Arkansas, Fayetteville, AR 72701
}
\author{Jana Shen} 
\affiliation{
Department of Pharmaceutical Sciences, University of Maryland School of Pharmacy, Baltimore, MD 21201
}
\email{jana.shen@rx.umaryland.edu}
\author{Guo-Wei Wei}
\affiliation{
Department of Mathematics \&  Department of Electrical and Computer Engineering \& Department of Biochemistry and Molecular Biology,
Michigan State University, MI, 48824, USA
}
\email{weig@msu.edu}

\begin{document}

\begin{abstract}
The continuous evolution of the SARS-CoV-2 virus poses a significant challenge to global public health. 
Of particular concern is the potential resistance to the widely prescribed drug PAXLOVID, of which the main ingredient nirmatrelvir inhibits the viral main protease (Mpro). 
Here, we developed CAPTURE (direCted flAg laPlacian Transformer for drUg Resistance prEdictions)
to analyze the effects of Mpro mutations
on nirmatrelvir-Mpro binding affinities
and identify potential drug-resistant mutations.
CAPTURE combines a comprehensive mutation analysis with a resistance prediction module based on
DFFormer-seq, which is a novel ensemble model that leverages a new Directed Flag Transformer and sequence embeddings from the protein and small-molecule-large-language models.
Our analysis of the evolution of Mpro mutations revealed a progressive increase in mutation frequencies for residues near the binding site between May and December 2022, suggesting that the widespread use of PAXLOVID created a selective pressure that accelerated the evolution of drug-resistant variants.
Applied to mutations at the nirmatrelvir-Mpro binding site,
CAPTURE identified several potential resistance mutations, including H172Y and F140L, which have been experimentally confirmed, as well as five other mutations that await experimental verification.
CAPTURE evaluation in a limited experimental data set on Mpro mutants gives a recall of 57\% and a precision of 71\% for predicting potential drug-resistant mutations.
Our work establishes a powerful new framework for predicting drug-resistant mutations
and real-time viral surveillance. 
The insights also guide the rational design of more resilient next-generation therapeutics.
\end{abstract}


\section{Introduction}

COVID-19 has entered an endemic phase; however, variants of SARS-CoV-2 continue to circulate throughout the world, posing a sustained challenge to public health.
The rapid and continuous emergence of new variants raises concerns about potential resistance to existing COVID-19 treatments \cite{heilmann2022sars,iketani2023multiple}, 
particularly given the widespread reliance on Pfizer's PAXLOVID \cite{murphy2022covid,iketani2023multiple}, 
which is currently the primary FDA-approved oral antiviral therapy. 
The main ingredient of PAXLOVID is nirmatrelvir, a small molecule that inhibits the viral main protease (Mpro), a critical enzyme for viral replication\cite{gao2021perspectives,xiong2022molecular}. 
Mutations in Mpro can potentially alter the structure and conformational dynamics of the protein, reducing the binding affinity (BA) and the efficacy of nirmatrelvir \cite{li2023therapeutic,hu2023naturally,clayton2023integrative}. 

\begin{figure*}[!ht]
\centering
\includegraphics[width=\textwidth]{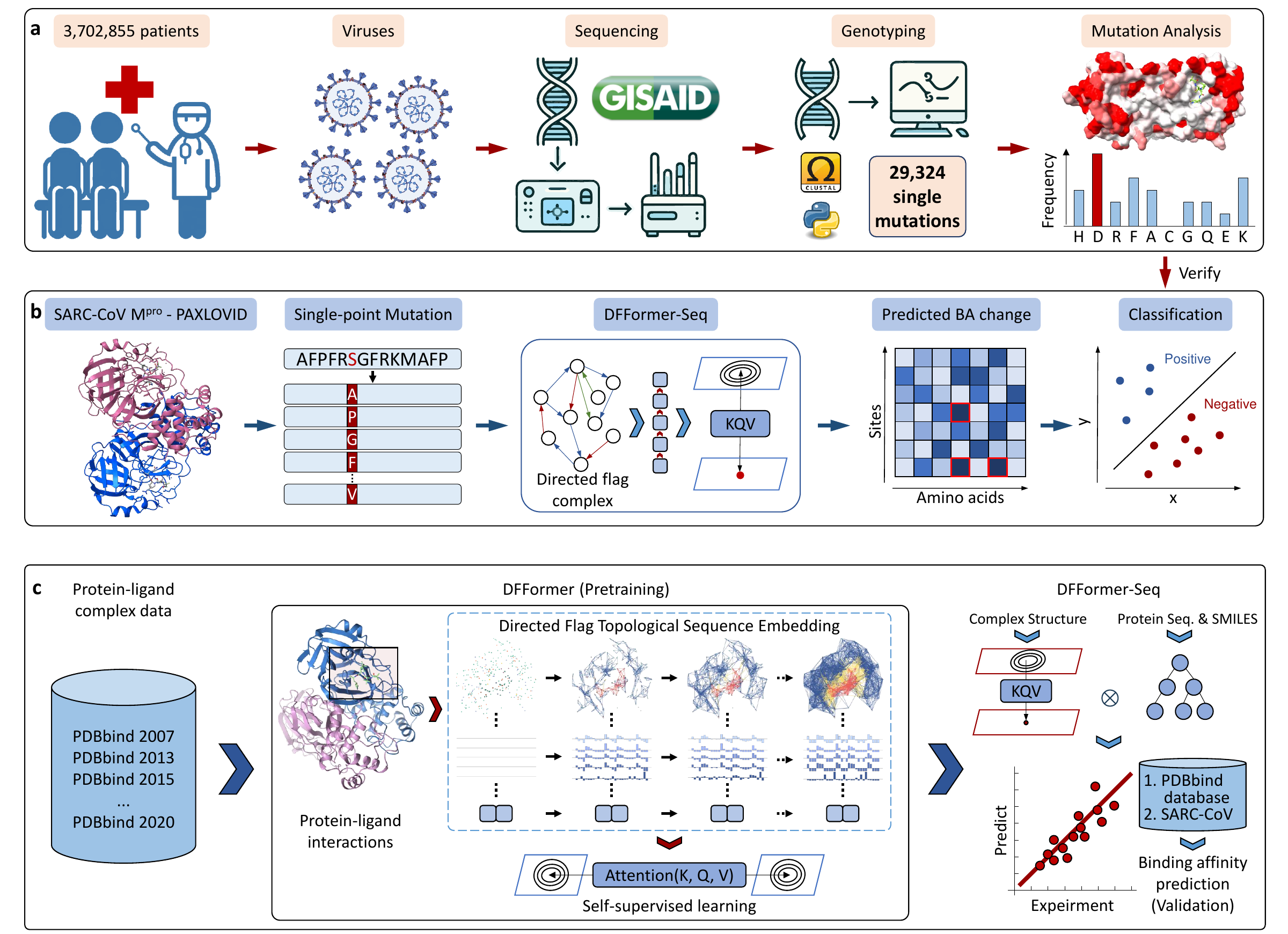}
\caption{\textbf{Overview of CAPTURE (direCted flAg laPlacian Transformer for drUg Resistance prEdictions).} 
\textbf{a}, The mutation data analysis module extracts viral sequences and frequencies of mutations at the drug target of interest from a database (e.g., GISAID). 
These mutations are evaluated for drug resistance potential in the prediction module.
\textbf{b}, The prediction module uses DFFormer-seq to analyze how mutations change drug binding affinities, and the results are classified for potential drug resistance.
\textbf{c}, DFFormer-seq is pretrained on a dataset of label-free protein-ligand complex structures followed
by a two-stage fine-tuning for predicting
ligand binding affinities. 
}
\label{Fig:CAPTURE}
\end{figure*}

Motivated by the need to forecast potential drug resistance mutations for Mpro and other drug targets\cite{yu2023deciphering,rimal2024saambe}, we developed a workflow CAPTURE (direCted flAg laPlacian Transformer for drUg Resistance prEdictions)
that performs mutation data analysis and drug resistance predictions (Fig.~\ref{Fig:CAPTURE}).
The mutation analysis module (Fig.~\ref{Fig:CAPTURE}a) retrieves and processes the genomic sequences and mutation frequencies related to the drug target of interest from the database GISAID 
(Global Initiative on Sharing All Influenza Data) \cite{shu2017gisaid}, which tracks, among other influenza viruses, variants of SARS-CoV-2 (also known as hCoV-19).
The prediction module (Fig.~\ref{Fig:CAPTURE}b) makes use of an ensemble deep learning model (DFFormer-seq) 
comprised of
Directed Flag transformers (DFFormer, Fig.~\ref{Fig:CAPTURE}c) and sequence embedding decision trees
to predict mutation-induced binding affinity changes,
which are then processed for drug resistance classification.

The transformer architecture \cite{vaswani2017attention} has been used to train Large Language Models (LLMs) such as ChatGPT \cite{brown2020language, ouyang2022training}, which leverage extensive pretraining and strategic use of unlabeled data. 
These models offer enormous potential for self-supervised machine learning (ML) in domains where traditional labeled data are scarce, expensive, or inadequate \cite{vaswani2017attention,ouyang2022training}, for example,
computational drug discovery \cite{chen2021algebraic,li2021effective,li2021machine}. 
However, a major challenge in applying LLM-inspired models to drug discovery is that these models are primarily designed for sequential data, such as text, and cannot handle 3D information such as the structures of protein-ligand complexes. 
However, spatial arrangement is important for developing models that accurately predict drug-binding affinities and selectivities.

DFFormer addresses
the aforementioned challenge by converting the complex 3D structural information of protein-ligand complex into a 1D sequential representation through the Persistent Directed Flag Laplacian theory \cite{jones2024persistent}, also know as the directed clique Laplacian theory, 
making it suitable for developing LLM-inspired deep-learning models. 
The 3D-to-1D transformation is achieved through directed flag Laplacian techniques that capture the intricate geometric and topological features of the protein-ligand binding site. 
The integration of multiscale-directed flag Laplacian analysis and deep learning enables nuanced learning and prediction of biomolecular interactions. 

We applied CAPTURE to analyze and predict drug-resistant mutations in SARS-CoV-2 Mpro.
Analysis of clinical data on Mpro mutations from five time periods between May and December 2022 reveals that the mutation frequencies of residues located within 15 {\AA} of nirmatrelvir increase consistently over the five periods.
Moreover, a dramatic increase is observed in the period of October-December 2022, suggesting that the widespread use of PAXLOVID has promoted mutations close to the inhibitor binding site, which may substantially increase the risk of
drug resistance.

To identify potential drug resistance mutations of Mpro, we applied DFFormer-seq to analyze the effects of Mpro binding site mutations on the affinities of nirmatrelvir followed by drug resistance classification.
DFFormer-seq was trained on the PDBbind 2020 dataset \cite{liu2015pdb,cheng2009comparative,li2014comparative,su2018comparative} and offers the state-of-the-art prediction accuracy for protein-ligand binding affinities based on commonly used benchmarks, CASF-2007, CASF-2013, and CASF-2016.
\cite{cheng2009comparative,li2014comparative,su2018comparative}.
%
In particular, recent experimental studies suggested that two predicted mutations are resistant to nirmatrelvir. \cite{hu2023naturally,iketani2023multiple}
These results demonstrate that CAPTURE establishes a powerful computational framework to systematically
predict and monitor the impact of emerging variants of SARS-CoV-2 and other viruses on drug resistance,
informing the proactive design of resilient next-generation therapeutics. 
Beyond deep mutation analysis, 
DFFormer-seq has broad utility in early-stage drug discovery, where protein-ligand affinities drive molecular recognition and therapeutic efficacy.

\section{Results}

\paragraph{Overview of the CAPTURE workflow.}
We developed a workflow CAPTURE (direCted flAg laPlacian Transformer for drUg Resistance prEdictions) to assess the impact of mutations on drug resistance given a three-dimensional structure of a protein-ligand complex (Fig.~ \ref{Fig:CAPTURE}).
CAPTURE consists of the mutational data analysis and resistance prediction modules.
The analysis module performs sequence analysis and retrieves mutation data for a specific drug target (Fig.~\ref{Fig:CAPTURE}a). 
For SARS-CoV-2 Mpro, we used the GISAID database   
\cite{shu2017gisaid}, which tracks the variants of SARS-CoV-2 and other influenza viruses.
Single nucleotide polymorphism (SNP) was used to study the genotypic changes of SARS-CoV-2. 
By aligning the sequences with the reference genome, we collected 3,702,855 SNP profiles from complete genomes by May 1, 2024. After removing duplicates, we identified 29,324 mutations, of which 1,158 are in open reading frame 1 (ORF1) that encodes the Mpro. 

\begin{figure*}[!ht]
\centering
 \includegraphics[width=\textwidth]{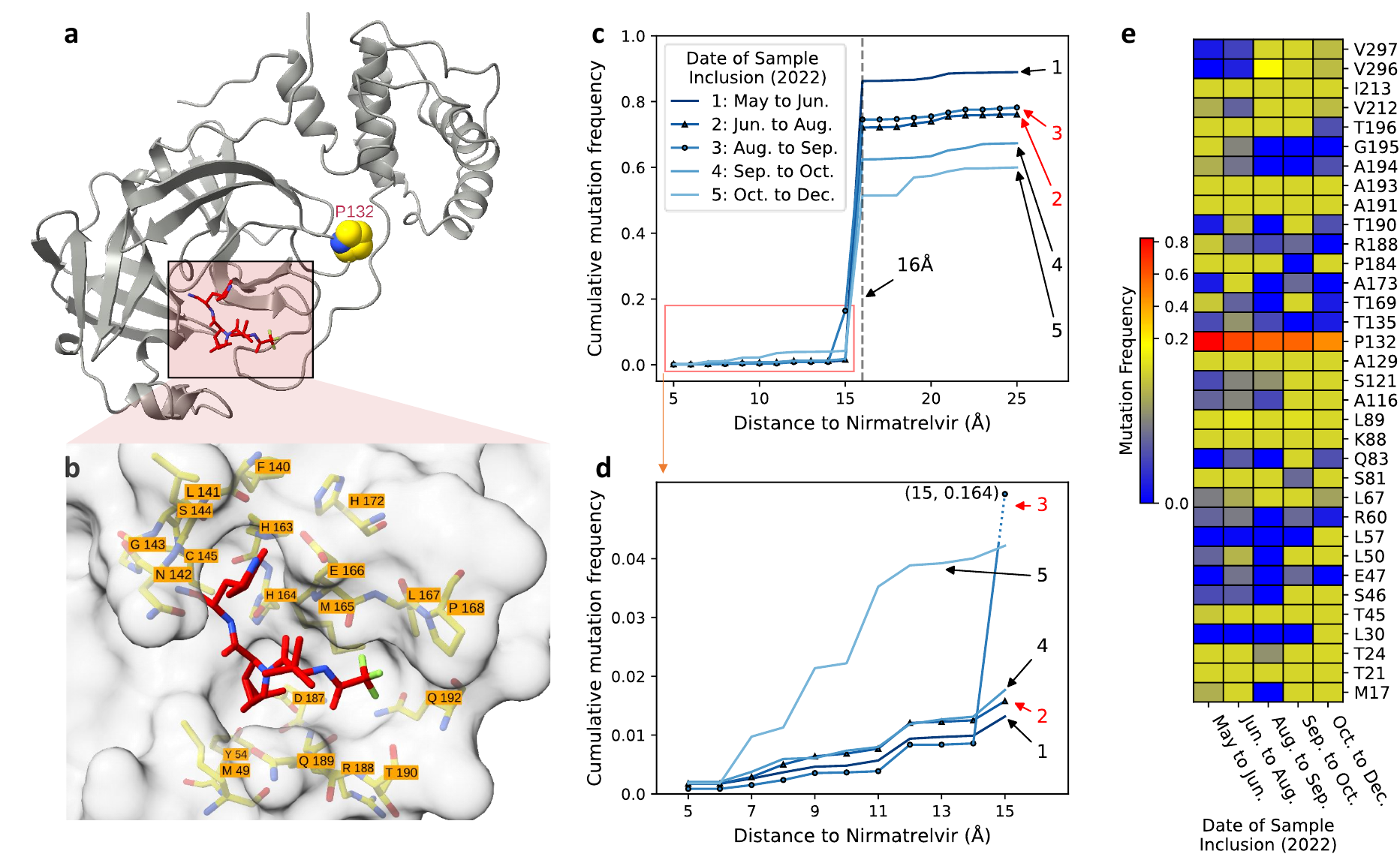}
\caption{\textbf{Analysis of SARS-CoV-2 Mpro mutations suggests an increasing risk of nirmatrelvir resistance due to mutations near the binding site.} 
\textbf{a}, The X-ray co-crystal structure of SARS-CoV-2 Mpro in complex with nirmatrelvir (PDB ID: 7VH8 \cite{zhao2022crystal}). 
For clarity, only one protomer is shown in the cartoon, 
with nirmatrelvir rendered in the stick model.
The mutation site of the Omicron variant, P132, is highlighted in the sphere model.
\textbf{b}, A close-up view of the Mpro binding site. 21 residues forming interactions with nirmatrelvir are labeled.
\textbf{c}, Cumulated mutation frequency of the Mpro residues within varying distances from nirmatrelvir. The latter refers to the minimal distance
between any heavy atoms of the mutated residue and nirmatrelvir.
Each data point represents the mutation frequency of all residues within a specified distance from the inhibitor during a time period normalized by the number of tested samples in the period.
Five time periods from May to December 2022 are shown 
(generated from GISAID \cite{shu2017gisaid}).
\textbf{d}, A zoomed-in view of the cumulative mutation frequencies with a 15-{\AA} radius from nirmatrelvir.
\textbf{e}, Mutation frequencies of individual 34 residues within 16 {\AA} from nirmatrelvir. 
Residues with absolute frequencies below 10 are not shown.
}
\label{Fig:mutation_analysis}
\end{figure*}

The prediction module uses DFLFormer-seq to predict BA changes due to mutations, followed by a binary classification to identify potential drug resistance mutations (Fig.~\ref{Fig:CAPTURE}b).
In this work, we restricted ourselves to the binding site residues of Mpro as well as P132, which is the mutation site of the Omicron variant (Fig.~\ref{Fig:mutation_analysis}a,b).
For each residue, 19 possible mutations are considered.
For each mutation, we retrieved the X-ray structure of the mutant Mpro-ligand complex from the PDB or mutated the residue using Jackal\cite{xiang2001extending} based on the structure of the WT Mpro-nirmatrelvir complex (PDB ID 7VH8) as a template if an experimental structure of the mutant-ligand complex is not available.
The structure underwent a brief energy minimization using Jackal and Schr\"{o}dinger's optimization tool. 
The trained task-specific DFFormer-seq model was then used to predict the BA. 
Based on the predicted BAs of the mutant complexes, a binary classification was performed to identify potential drug resistance mutations (positives).

\paragraph{Analysis of the Mpro mutation data suggests an increasing risk of drug resistance.}
To inform about the potential impact of PAXLOVID usage on SARS-CoV-2 evolution, we analyzed the evolution of Mpro residues, comparing periods before and after widespread use of PAXLOVID. 
Fig.~\ref{Fig:mutation_analysis}c displays the mutation frequencies of residues within varying distances from nirmatrelvir over five periods between May and December 2022.
Mutation frequencies are extremely low for residues located within 15 {\AA} of the inhibitor,
which is expected, as most residues near the inhibitor belong to the active site of the protein and are conserved or highly conserved.
The mutation frequencies of the five periods show a sharp increase of nearly two orders of magnitude around 16 {\AA} from the inhibitor, which
is due to the Omicron mutation at P132 (the minimal distance to nirmatrelvir is 15.8 {\AA}).
Interestingly, while the overall Mpro mutation frequencies decrease over time from May to December 2022, as seen from the curves outside of 16 {\AA} (Fig.~\ref{Fig:mutation_analysis}c),
a nearly reversed trend is observed for residues within 15 {\AA} of the inhibitor, that is, each period exhibits higher mutation frequencies than the preceding period (Fig.~\ref{Fig:mutation_analysis}d). 
Remarkably, residues within 7 {\AA} of nirmatrelvir
show an order of magnitude increase in mutation frequencies in the period of October to December 2022 compared to all previous periods
(Fig.~\ref{Fig:mutation_analysis}d).
From October to December 2022, the mutation frequencies increased even more dramatically for residues located within 15 {\AA} of the inhibitor but outside its immediate binding site, compared to all previous periods.
The trend of elevated mutation frequencies with 15 {\AA}
from the inhibitor correlates with PAXLOVID usage patterns, which peaked in July 2022 (surveyed January- September 2022) \cite{murphy2022covid}  and likely again during the December Omicron surge, compared to earlier months. 
Together, these data suggest that the widespread usage of PAXLOVID has promoted mutations close to the inhibitor binding site, which may substantially increase the risk of drug resistance.
 
To examine the mutational evolution of individual residues within 16 {\AA} from nirmatrelvir for the five periods mentioned above, a heat map is displayed in Fig.~\ref{Fig:mutation_analysis}e.
Residue P132, which is mutated to histidine (P132H) in the SARS-CoV-2 Omicron variant \cite{sacco2022p132h,
Ibrahim_Hilgenfeld_2024_hLife} displays the highest mutation frequency in the five periods compared to other sites; 
however, the mutation frequency decreases over time from 0.82 in May-June to 0.46 in October-December 2022.
This trend may reflect the decreasing prevalence of the Omicron variant as newer SARS-CoV-2 variants emerged and became more dominant in the population.
V296 (located 15 {\AA} from the inhibitor) is another hot spot for mutation. Its mutation frequency peaks at 0.16 from August to September 2022, which is responsible for the peak shown in Fig.~\ref{Fig:mutation_analysis}d. 
The heat map (Fig.~\ref{Fig:mutation_analysis}e)
also reveals residues that exhibit a sharp increase in mutation frequencies in the period of October to December compared to the preceding time periods, e.g.,
S121, A116, S81, L57, L50, S46, L30, which
contribute to the dramatically increased mutation frequencies from October to December as shown in Fig.~\ref{Fig:mutation_analysis}d.

\begin{figure*}[!ht]
\centering
\includegraphics[width=\textwidth]{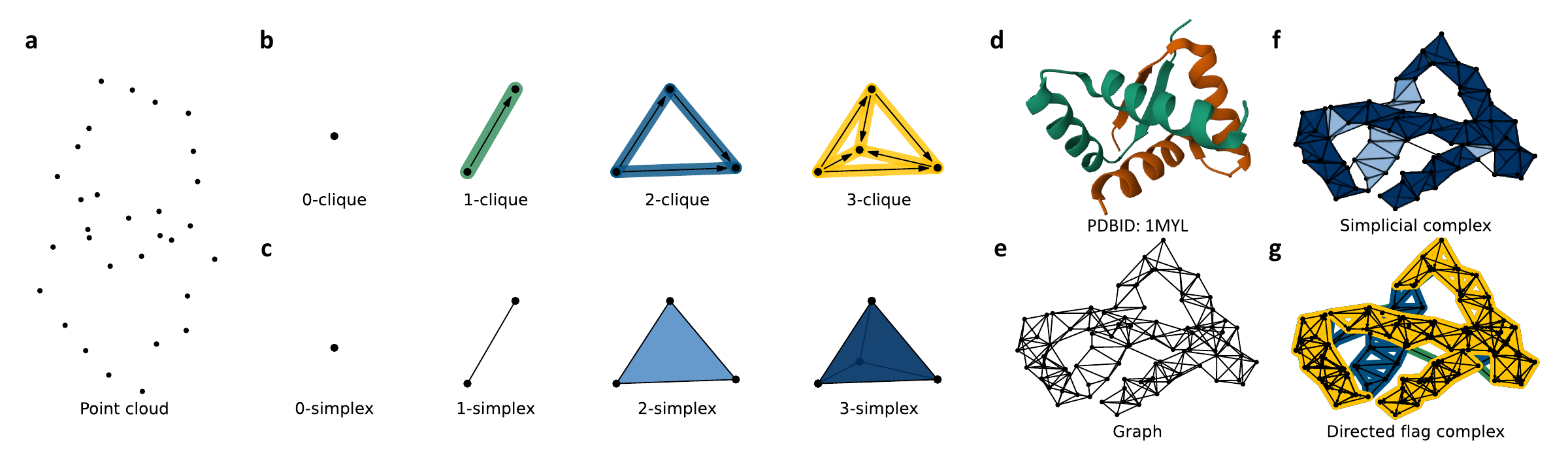}
\caption{\textbf{Schematics of the directed flag complex concept.}
\textbf{a}, Point cloud representations of structural data.
\textbf{b}, Basic elements of a directed flag complex:
0-clique, 1-clique, 2-clique, and 3-clique.
\textbf{c}, Components of a simplicial complex:
0-simplex (node), 1-simplex (edge), 2-simplex (triangle), and 3-simplex (tetrahedron).
\textbf{d}, Visualization of a protein structure (PDB 1MYL).
\textbf{e}, Graph representation of the protein C$\alpha$ atoms (1MYL).
\textbf{f, g}, Simplicial (f) and directed flag (g) complex representation of the protein C$_\alpha$ atoms.
The 2- and 3-simplices are shown in light and dark blue, respectively (f).
The 1-, 2-, and 3-cliques are shown with arrows on a green, blue, and yellow background, respectively (g).
    }
    \label{Fig:directed_flag_topology}
  \end{figure*}

\paragraph{DFFormer-seq: an ensemble predictor.}
Inspiring by TopoFormer\cite{chen2024multiscale}, which utilizes a hyperdigraph\cite{chen2023persistent} representation of protein-ligand complexes, we developed DFFormer, which leverages the transformer architecture with an element-specific Directed Flag Laplacian method for topological sequence embeddings
(Fig.~\ref{Fig:directed_flag_topology} and Supplementary Protocol Fig. S1).
As model pretraining, self-supervised learning on the X-ray structures of 19,513 unique protein-ligand complexes from PDBbind 2020 \cite{liu2015pdb,cheng2009comparative,li2014comparative,su2018comparative}
taken from the Protein Data Bank (PDB) was conducted, whereby the loss function is the reconstruction of the sequences.
To fine-tune DFFormer for the task of structure-based BA prediction, 18,904 protein-ligand BAs from PDBbind 2020 (excluding CASF core sets) \cite{liu2015pdb,cheng2009comparative,li2014comparative,su2018comparative} were used.
20 instances of DFFormer were trained, each initialized with a distinct random seed. 

To further enhance the model's predictive power, we incorporated sequence-based models. 
Specifically, we used the large protein language model ESM \cite{rives2021biological} to extract features of protein sequences
and the transformer-CPZ model \cite{chen2021extracting} to extract features of SMILES from the aforementioned protein-ligand complexes. 
Using the features of protein and small molecule sequences, 
we trained 20 gradient-boosting decision tree models, referred to as Seq-trees. 
Subsequently, a final prediction model DFFormer-seq was built,
which is an ensemble of 10 randomly
selected models from 20 DFFormers and 20 Seq-trees.

\begin{figure*}[!ht]
\centering
\includegraphics[width=5.5in]{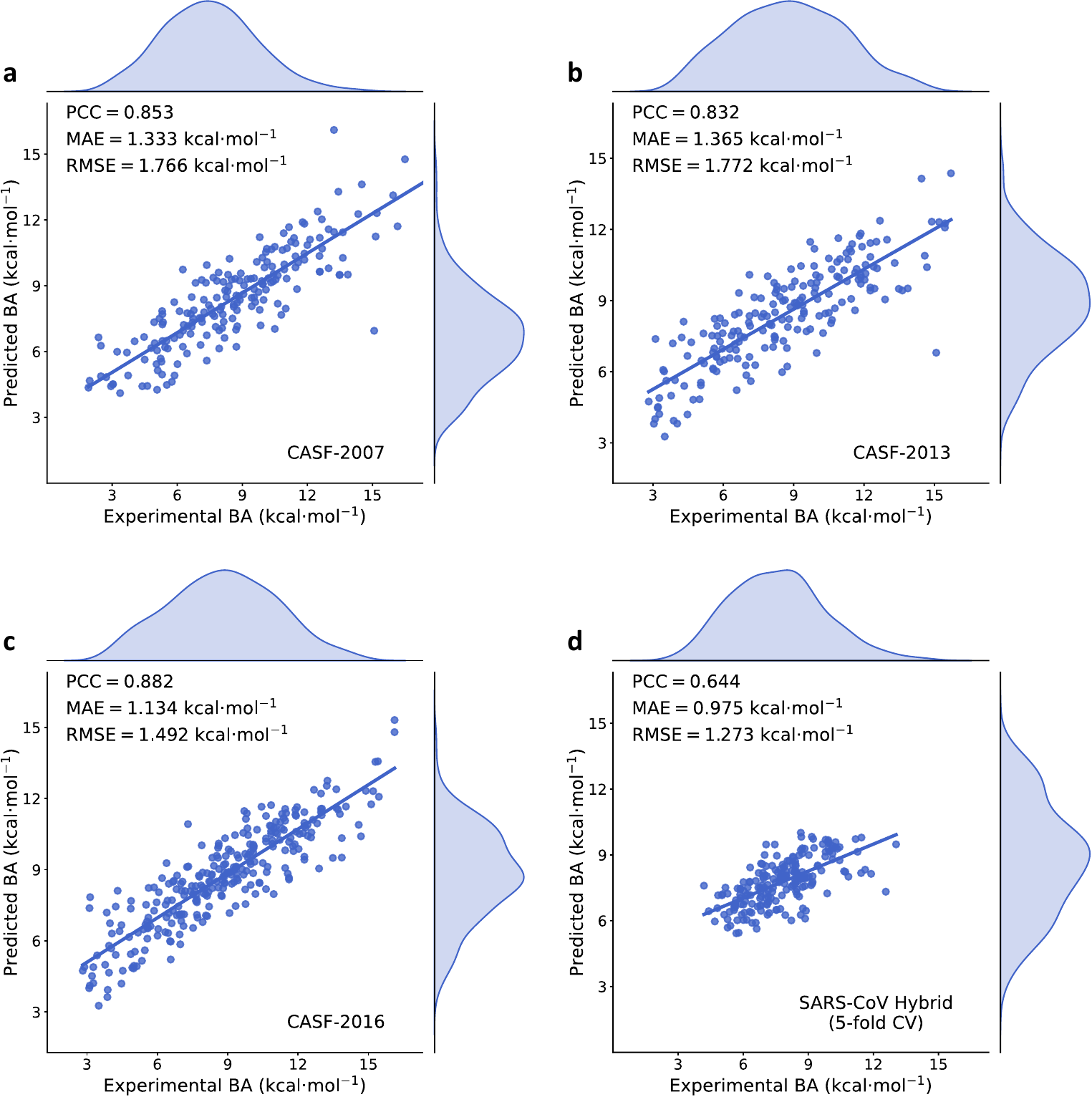}
\caption{\textbf{Performance of DFFormer-seq in predicting protein-ligand binding affinities (BAs).} 
Comparison between the experimental and predicted BAs for the 
test sets, CASF-2007 (\textbf{a}), CASF-2013 (\textbf{b}), 
and CASF-2016 (\textbf{c}), and SARS-CoV/CoV-2 Mpro (\textbf{d}).
The Pearson's correlation coefficient (PCC), 
mean average error (MAE), and root-mean-square error (RMSE) are shown. 
}
\label{Fig:BA_tests}
\end{figure*}

\begin{table*}[!htb]
\centering
\caption{Comparison of PCC/RMSE between DFFormer-seq and 11 published 
models on CASF benchmark datasets$^{a}$
}
\resizebox{4.5in}{!}
{
\begin{tabular}{lllll}
\hline
\hline
Model & Training set & CASF-2007 & CASF-2013 & CASF-2016\\
       & PDBbind     & n=195 & n=195 & n=285 or 290$^{b}$\\
\hline
       Pafnucy \cite{stepniewska2018development} & 2016 (11,906)   &                    &                    & 0.78               \\ 
       Deep Fusion \cite{jones2021improved}      & 2016 (9,226)    &                    &                    & 0.803/1.809$^{b}$ \\ 
       graphDelta \cite{karlov2020graphdelta}    & 2018 (8,766)    &                    &                    & 0.87/\textbf{1.431}$^{b}$   \\ 
       DeepAtom \cite{rezaei2020deep}            & 2018 (9,383)    &                    &                    & 0.831/1.680$^{a}$ \\ 
       SE-OnionNet \cite{wang2021se}             & 2018 (11,663)   &                    & 0.812/2.307       & 0.83               \\ 
       ECIF \cite{sanchez2021extended}           & 2019 (9,299)    &                    &                    & 0.866/1.594       \\ 
       OnionNet-2 \cite{wang2021onionnet}        & 2019 ($>$9,000) &                    & 0.821/1.850       & 0.864/1.587       \\ 
       PLANET \cite{zhang2023planet}             & 2020 (15,616)   &                    &                    & 0.824/1.700$^{b}$ \\ 
       HydraScreen \cite{prat2023hydrascreen}    & 2020$^{c}$    &                    &                    & 0.86/1.568 \\
       Point VS \cite{scantlebury2023small}      & 2020 (19,157)   &                    &                    & 0.816 \\
       HAC-Net \cite{kyro2023hac}                & 2020 (18,818)   &                    &                    & 0.846/1.643 \\ 
       DFFormer-seq                              & 2020 (18,904)   &  \textbf{0.853}/\textbf{1.766} & \textbf{0.832}/\textbf{1.772} & \textbf{0.882}/\textbf{1.492}       \\
       &           &      &        &\textbf{0.884}/1.479$^{b}$ \\
  \hline
    \hline
       \end{tabular}
}
   \label{Tab:compare_models}
   \begin{minipage}{\textwidth}
   \footnotesize
$^{a}$The best PCC and RMSE among all methods are in bold font. 
For the training set, the (year) version of PDBbind \cite{liu2015pdb,cheng2009comparative,li2014comparative,su2018comparative} and sample size (in parenthesis) are given.
$^{b}$The CASF-2016 dataset plus five additional 
protein-ligand complexes (also known as the PDBbind 2016 core set). 
$^{c}$The training data size not mentioned in Ref \cite{prat2023hydrascreen}. 
\end{minipage}
  \end{table*}

\paragraph{Evaluation of DFFormer-seq for Predicting Protein-Ligand BAs.}
We evaluated the performance of DFFormer-seq using the common benchmark datasets, CASF-2007, CASF-2013, and CASF-2016.
The model showed consistent performance across these benchmark sets, 
with the respective Pearson's correlation coefficients (PCCs) of 0.853, 0.832, and 0.882, and root-mean-square errors (RMSEs) of 1.766, 1.772, and 1.492 kcal/mol (Fig.~\ref{Fig:BA_tests}a-c).
The performance metrics of the separate Seq-trees and DFFormer models are given in Supplementary Data Fig. S1. 
Table~\ref{Tab:compare_models} compares the PCC and RMSE of DFFormer-seq with 11 other published ML models
(Refs \cite{stepniewska2018development,jones2021improved,karlov2020graphdelta,rezaei2020deep,wang2021se,sanchez2021extended,wang2021onionnet,zhang2023planet,prat2023hydrascreen,scantlebury2023small,kyro2023hac}) 
based on CASF benchmarks.
Note that 7 of the 11 published models were trained on the significantly smaller PDBbind 2019 dataset, while 4 of them were trained on the same PDBbind 2020 dataset.
Although the goal here is to build the most powerful model for drug resistance application using the 2020 general set, we also provide details of the models trained on the smaller refined set\cite{li2021machine}, which is a subset of the general set, in the Supplementary Notes section for a comprehensive review.
DFFormer-seq gives the highest PCCs across all three benchmark sets
and the lowest RMSEs for the CASF-2007 and CASF-2013 datasets,
while the RMSE (1.492 kcal/mol) of CASF-2016 is slightly lower than that (1.431 kcal/mol) reported for the graphDelta model. \cite{karlov2020graphdelta} 
However, it should be noted that our predictions represent an ensemble of 20 models, whereas the evaluation of graphDelta was for a single model selected at 2000 epochs, which may be less robust due to model variability. 
These results suggest that DFFormer-seq delivers state-of-the-art performance in predicting protein-ligand BAs. 

\paragraph{Further fine-tuning DFFormer-seq for predicting resistance mutations in Mpro.}
To deploy DFFormer-seq for predicting the potential drug resistance mutations of Mpro, we further fine-tuned it on the SARS-CoV-2/CoV Mpro dataset, which contains the BAs of 203 inhibitors (excluding nirmatrelvir) against SARS-CoV-2 or SARS-CoV Mpro (Table~\ref{table:dataset_info}, Supplementary Data Table S1 and S2).
Specifically, a five-fold cross-validation of the DFFormer-seq was performed on this dataset (Supplementary Data Table S3) before
it was finalized using the entire SARS-CoV-2/CoV dataset. 

The average PCC and RMSE across five folds are 0.722 and 1.437 kcal/mol, respectively (Supplementary Data Table S3).
Compared to the CASF benchmarks, the RMSE is significantly lower, by 0.2 kcal/mol compared to 
CASF-2016 and by 0.5 kcal/mol compared to CASF-2007 and CASF-2013 benchmarks (Fig.~\ref{Fig:BA_tests}\textbf{d}). 
However, the PCC is lower by nearly 0.2 as well, which can be attributed to the significantly narrower range of BAs (by about 4 kcal/mol, Fig.~\ref{Fig:BA_tests}\textbf{d}).
Given the extremely small dataset, these performance metrics are strong, suggesting that the pre-trained knowledge is effectively leveraged and the model has learned meaningful patterns and relationships within the data.
The performance of separate DFFormer and Seq-trees is given in Supplementary Data Fig. S1.

\paragraph{Evaluating the effect of binding-site mutations on nirmatrelvir-Mpro binding.}
We next applied DFFormer-seq fine-tuned on the Mpro data to predict the BA changes of mutants relative to the wild type (WT) for nirmatrelvir.
The co-crystal structure of the WT Mpro-nirmatrelvir complex \cite{zhao2022crystal} shows that 21 residues directly interact with nirmatrelvir (minimum distance to nirmatrelvir below 4.5 {\AA}), including Q192, T190, Q189, E166, R188, D187, Y54, M49, H164, M165, H41, C145, G143, S144, N142, L141, F140, H172, H163, E166, L167, and P168 (Figure~\ref{Fig:mutation_analysis}b). 
We focus on 19 of them, excluding H41 and C145, which are absolutely conserved catalytic residues in Mpro.
Note, we did not consider residues outside the inhibitor binding site because they have a much lower chance of inducing drug resistance, except for P132, which is the mutation site of the Omicron variant P132H.

\begin{figure*}[!ht]
\centering
\includegraphics[width=\textwidth]{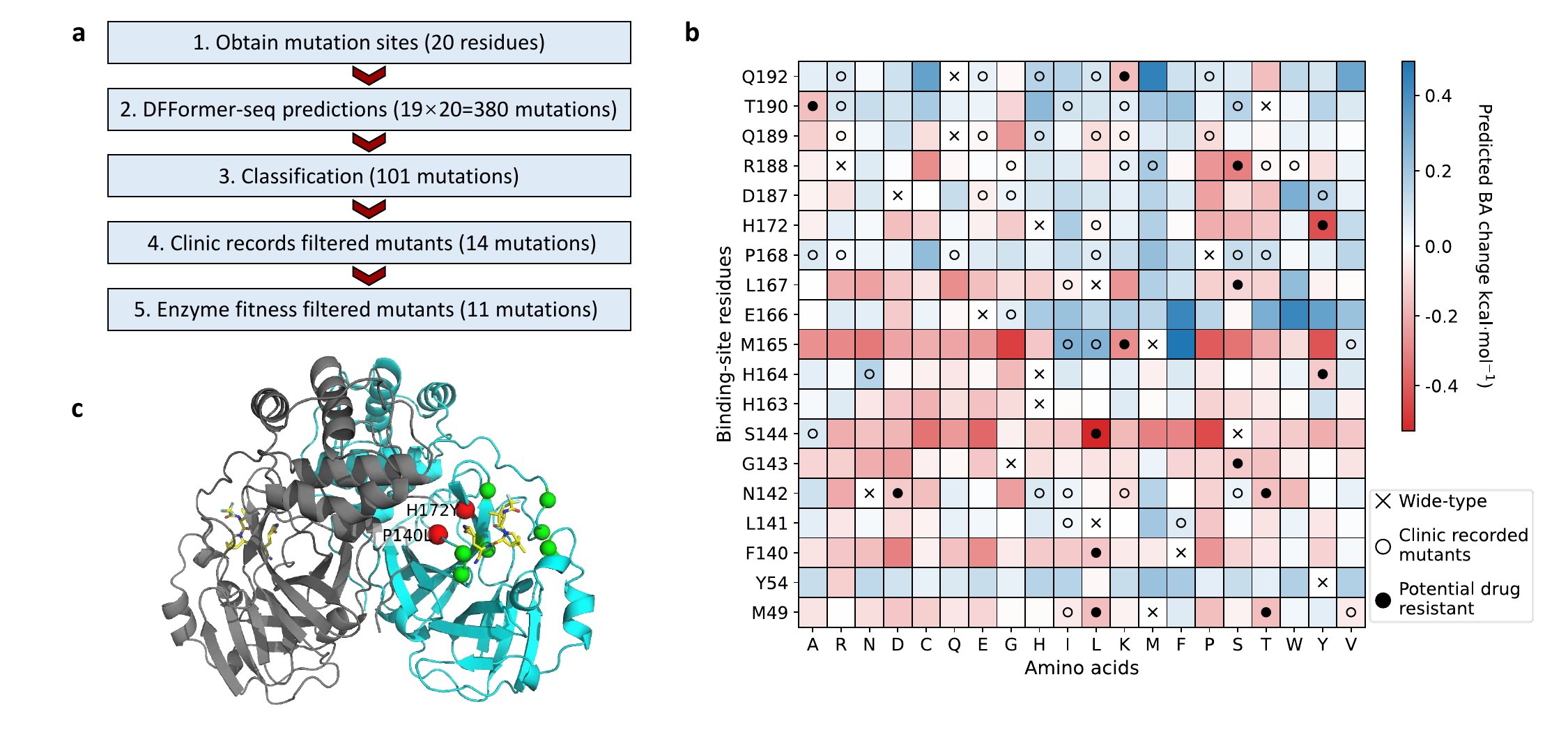}
\caption{\textbf{Prediction of drug-resistance mutations in SARS-CoV-2 Mpro.}
\textbf{a}, A flow chart for predicting drug resistance mutations in SARS-CoV-2 Mpro.
\textbf{b}, A heat map showing the predicted BA changes of nirmatrelvir for the single-point mutations ($x$-axis) at 19 binding-site residues ($y$-axis). Blue and red colors indicate increased and decreased BAs, respectively. X denotes the WT residues; circles identify natural mutations (according to GISAID) \cite{shu2017gisaid}, and filled circles indicate natural mutations predicted to be potentially drug-resistant.
\textbf{c,} Locations of predicted drug-resistance mutation sites in the SARS-CoV-2 Mpro structure. Nirmatrelvir is shown in the yellow stick model. The C$\alpha$ atoms of potential drug-resistance mutation sites are shown in spheres.
F140L and H172Y (red) are experimentally confirmed, 
while other predicted mutations G143S, H164Y, L167S, R188S, and T190A (green) are to be experimentally verified.
}
\label{Fig:Mpro_analysis}
  \end{figure*}

A flow chart for predicting drug-resistant mutations of Mpro is shown in Fig.~\ref{Fig:Mpro_analysis}a. 
The 19 possible mutations were considered for each of the 20 sites, resulting in a total of 380 mutations.
DFFormer-seq was used to predict the BA changes of the mutants relative to the WT.
The predicted $\Delta$BA values ($\Delta$BA=BA$^{\rm mut}$ - BA$^{\rm WT}$) are displayed on a heat map (Fig.~\ref{Fig:Mpro_analysis}b)
and also given in the SI (Table S4).

Recently, Hu et al. \cite{hu2023naturally} measured the
Mpro enzymatic activity ($k_\text{cat}$/$K_\text{m}$), thermal stability, and nirmatrelvir binding (\Ki) or the inhibition constant (\IC) for 97 naturally occurring mutations at 12 binding-site residues,
including H41, M49, T135, N142, S144, H163, H164, M165, E166, H172, Q189, and Q192.
Since our model was trained on mixed \Ki\ and \IC\ values and these data were obtained with varying experimental conditions, it is not meaningful to directly compare the predicted BA changes with the experimental data of Hu et al. \cite{hu2023naturally}
Instead, we treated the problem as a binary classification.
As suggested by
Hu et al. \cite{hu2023naturally}, 
a mutation is labeled positive (i.e., potentially drug resistant), if \Ki\ (or IC50 if \Ki\ is unavailable) shows at least a 10-fold increase compared to the WT, i.e., $\Delta$p\Ki$>$1;
otherwise, the mutation is labeled negative (Supplementary Data Table S5).

We evaluated the true and false positive prediction rates (TPR and FPR) by the DFFormer-seq model using different thresholds of $\Delta$BA to define the predicted positives. The receiver operating characteristic curve (ROC, Figure~\ref{Fig:Mpro_analysis}b) shows that the area under ROC (AUROC) is 0.60, suggesting that the model has a statistically higher chance of identifying a potential drug resistance mutation than a random guess. 
Youden's $J$ statistics, defined as $J =$ TPR - FPR, was used to identify a $\Delta$BA threshold of -0.09 kcal/mol to classify drug resistance predictions from DFFormer-Seq. 
Accordingly, a mutation with $\Delta$BA $<-0.09$ kcal/mol is predicted positive or potentially drug resistant.
At this threshold, the model achieves a recall (TPR) of 0.57, a precision of 0.71, and an accuracy of 0.77, with reference to the 97 mutations characterized experimentally. \cite{hu2023naturally}


\paragraph{Predicting drug-resistant mutations in Mpro.}
Next, the results of the mutation data analysis module were used to filter out clinically irrelevant mutations, i.e., those that have not been identified in clinical records.
A total of 59 clinically identified mutations were considered. 
This selection process resulted in 14 potentially drug-resistant mutations (Fig. \ref{Fig:Mpro_analysis}\textbf{b}, filled circles, and Supplementary Data Table S6). 
Interestingly, Omicron mutation P132H is among the predicted negatives ($\Delta$BA of 2.2 $\times$ 10$^{-2}$ kcal/mol), which is consistent with the experimental findings that the P132H mutation does not affect the potency of nirmatrelvir as well as two other Mpro inhibitors, despite the stability decrease of Mpro.
\cite{sacco2022p132h,Ibrahim_Hilgenfeld_2024_hLife}
Three of the predicted positive mutations, including S144L, M165K, and Q192K, were excluded as biologically unfeasible, due to a significant reduction in enzyme activity (loss of viral replicate fitness), defined by a $>$10-fold decrease in the $k_\text{cat}/K_\text{m}$ value relative to the WT Mpro. \cite{hu2023naturally}
Although four of the predicted positive mutations, 
M49L, M49T, N142D, N142T, are false positives,
which show negligible changes in the measured values of
{\Ki} or {\IC} (Supplementary Data Table S5), the remaining
seven are either consistent with the experiment
or must be experimentally verified.
Remarkably, H172Y shows the largest decrease in BA
among all predicted positive mutations, which
corroborates the experimental data showing a
25- \cite{clayton2023integrative} or 100-times 
\cite{hu2023naturally} increased {\IC} value for nirmatrelvir compared to the WT Mpro.
The predicted positive mutation F140L is consistent
with a recent cell-based study that inferred resistance to nirmatrelvir in Huh7-ACE cells.
\cite{iketani2023multiple} 
Our model also predicted potential drug resistance mutations that have not been experimentally tested, including
G143S, H164Y, L167S, R188S, and T190A, 
which represent testable hypotheses.

\section{Discussion}
We introduced a novel transformer-based workflow called CAPTURE for the prediction of drug resistance mutations.
CAPTURE consists of a mutation data analysis and a prediction module. 
The analysis module utilizes an external data source (e.g., GISAID) to compile a real-time list of clinical mutations in a drug target of interest, e.g., SARS-CoV-2 Mpro. 
The compiled mutations are submitted to the prediction module for AI testing of mutation-induced BA changes and making the prediction of potentially drug-resistant mutations.
The prediction module leverages a novel ensemble model (DFFormer-seq) comprising the newly developed directed flag Laplacian transformer (DFFormer) and decision trees based on LLM embeddings of protein and ligand sequences.
We demonstrated that DFFormer-seq delivers state-of-the-art accuracies in the prediction of protein-ligand BAs when compared to the published AI models.



To demonstrate the utility of CAPTURE, we applied it to identify potential resistance to nirmatrelvir mutations in SARS-CoV-2 Mpro, focusing specifically on mutations occurring at the residues of the drug binding site, as well as the Omicron mutation site P132. 
Together, 380 mutations from 20 sites were computationally tested for potential drug resistance (positives)
and the results were compared with those derived from experimental {\Ki} or {\IC} values of 97 single mutations at 12 binding site residues in Mpro.
Our model achieves a recall of 57\%, a precision of 71\%, and an accuracy of 77\%.
This performance is encouraging, considering that the
model is not trained on mutation-induced BA changes. 
Moreover, the Omicron mutation P132H was predicted as negative, consistent with the experiment. \cite{sacco2022p132h,Ibrahim_Hilgenfeld_2024_hLife}
After eliminating positive mutations that are clinically irrelevant or biologically unviable, CAPTURE predicted 11 potential drug-resistant mutations, including H172Y and F140L according to experiment
\cite{hu2023naturally,iketani2023multiple}
and five other mutations that are testable hypotheses.

Another significant finding of this work is related to the evolution of Mpro mutations. Our analysis revealed a concerning pattern: residues near the binding site showed progressively increasing mutation frequencies, with a dramatic increase in the period of October-December 2022. This striking acceleration in mutation rates suggests that the widespread use of PAXLOVID may be creating selective pressure, potentially driving the evolution of nirmatrelvir-resistant variants. This finding has important implications for the long-term effectiveness of this crucial COVID-19 treatment.

The work exposes a weakness of the DFLFormer model.
The predicted changes in BAs significantly underestimate the experimental changes in the values of {\Ki} or {\IC}.
This is not surprising given that the model is not trained on mutation-induced changes in BAs.
This discrepancy suggests a need for a more sophisticated protein modeling technique (e.g., the most recent AlphaFold3 model) \cite{Abramson_Jumper_2024_Nature}
and/or perhaps the need to incorporate multiple structure representations.
Future work will address this and other limitations, such as the neglect of double mutants. 
Despite these limitations, our work establishes a powerful new framework for predicting drug resistance mutations by combining directed flag Laplacian representations with deep learning and real-time genomic surveillance. 
This integrated approach not only demonstrates promise for forecasting resistant variants but also provides valuable insights for guiding the rational design of more resilient next-generation therapeutics. 

\section{Methods}\label{section:methods}

\subsection{The DFFormer architecture and training process}

The DFFormer combines the transformer architecture \cite{vaswani2017attention} with the persistent directed flag Laplacian method \cite{jones2024persistent}. 
The directed flag topological sequence embedding module (
Fig.~\ref{Fig:directed_flag_topology} and Supplementary Methods Fig. S1) enables the conversion of a 3D structure to a sequence of topological invariants, homotopic shapes, and stereochemistry. 
First, a substructure of the protein-ligand complex
is constructed using atoms within 20 {\AA} of any heavy atom of the ligand.
Next, the 3D substructure is converted to a directed flag topological sequence using a multiscale approach akin to a filtration process in algebraic topology.
Specifically, the scale ranges from 0 to 10 \AA\ in 0.1 \AA\ increments, leading to a directed flag topological sequence of 100 units. 
In each filtration step, the embedded features are represented by a 143$\times$6 matrix, with 6 attributes for each ${\cal{L}}_{0}$. The output of directed flag topological embedding is a summation of these topological embeddings and trainable multiscale embeddings. 
To reshape the 143$\times$6 matrix into a 1-dimensional vector, the model incorporates a convolutional layer in both the encoder and decoder (Supplementary Methods Fig. S1). This transformation is achieved through the persistent directed flag Laplacians technique, resulting in a series of embedding vectors. More details are given in Supplementary Methods.

The next module in the DFFormer conducts self-supervised learning with unlabeled data using a typical encoder-decoder transformer architecture (Supplementary Methods Fig. S1).
Here we used the unlabeled protein-ligand complex structures from PDBbind 2000.
The directed flag topological embeddings of these complexes
are reconstructed to calculate the loss using mean square error (MAE) as the metric. 
This self-supervised learning method allows the model to develop deep and generalizable representations of protein-ligand complexes from a substantial volume of unlabeled data, potentially streamlining the subsequent fine-tuning process. 

Following pretraining, the next module conducts supervised learning with labeled data, 
i.e., 18,904 BAs of protein-ligand complexes from PDBbind 2020 that exclude the core sets of CASF-2007, CASF-2013, and CASF-2016.
As the second stage of model fine-tuning, the 5-fold cross-validation on the BAs of SARS-CoV/CoV-2 Mpro.
Throughout these fine-tuning tasks, the MSE was used as the loss function.
The settings and hyperparameters of the DFFormer are given in the Supplementary Protocols.

\subsection{Datasets}
Table \ref{table:dataset_info} summarizes the datasets used in this study. 
For model pretraining, 19,513 cocrystal structures of protein-ligand complexes from the combined datasets of
PDBbind 2020, CASF-2007, CASF-2013, CASF-2016 \cite{liu2015pdb} were used with the duplicates removed. 
For model fine-tuning, PDBbind 2020 with the excluded CASF core sets was used, which contains the cocrystal structures and BAs of 18,904 protein-ligand complexes.
For model testing, the CASF-2007, CASF-2013, and CASF-2016 datasets were used. 
For the second-stage model fine-tuning, we used
the data in Ref. \citet{nguyen2020unveiling},  which contains the molecular structures and BAs of 155 SARS-CoV-2 Mpro complexes (Supplementary Table S1). 
Additional molecular structures and BAs of 48 Mpro inhibitors sourced from PDB were included (Supplementary Table S2). 



\begin{table}[!ht]
\centering
\caption{Summary of datasets used in this work}
\resizebox{1\columnwidth}{!}{
\begin{tabular}{lll}
 \hline
 \hline
\textbf{Dataset} &
\textbf{Size} &
\textbf{Usage and description} \\ 
\hline
Combined  &
          19,513 &
 Structures for   \\
PDBbind \cite{liu2015pdb,cheng2009comparative,li2014comparative,su2018comparative}& & pretraining; PDBbind 2020 \\ 
& & with duplicates removed\\
\hline
PDBind 2020 \cite{liu2015pdb,su2018comparative} & 18,904 &
For model fine-tuning;  \\ 
& & the CASF core sets excluded \\
\hline
CASF-2007;  & 675/680 &
    Hold-out for model testing; \\
 -2013; -2016\cite{cheng2009comparative,liu2015pdb}   & & the CASF core sets \\
\hline
SARS-CoV-2/CoV &
          203 &
For further fine-tuning;  \\
Mpro (Table S1, S2)& & Molecular structures and BAs\\
Mpro mutants  & 380 & for prospective analysis; \\
& & binding-site, P132 mutations\\ 
\hline
\hline
      \end{tabular}
    
    \label{table:dataset_info}
    }
  \end{table}

For the analysis of SARS-CoV-2 Mpro mutations,
we collected and processed the data following three steps.
First, we downloaded the SARS-CoV-2 genome sequences from the GISAID database \cite{shu2017gisaid} (\url{https://www.gisaid.org/}).
Next, we filtered out incomplete genome sequences
or those that do not have submission dates. 
We then aligned the complete genome sequences with the reference SARS-CoV-2 genome using Clustal Omega \cite{sievers2014clustal}, using the default settings for multiple sequence alignment. 
Our dataset contains 3,694,942 complete SARS-CoV-2 genome sequences as of November 29, 2023. 
To investigate the evolution of Mpro mutations and their potential impact on drug resistance, we included genome sequences from key dates throughout 2022, i.e.,  May 10, June 10, August 30, September 30, October 31, and December 31.

\section*{Data availability}
The training and benchmark data used in the work can be downloaded from the PDBbind website: http://www.pdbbind.org.cn/index.php. 
Sequence-based features and the SARS-CoV/CoV-2 Mpro dataset 
are freely downloadable at https://github.com/WeilabMSU/CAPTURE. 

\section*{Supporting Information}
Supporting Information is contained in three files: Supplementary Computational Methods, Protocols,
Supplementary Data, and Supplementary Mathematical Methods.

\section*{Code availability}
All source codes and models are freely available at https://github.com/WeilabMSU/CAPTURE.

\section*{Acknowledgments}
This work was supported in part by NIH grants R01AI164266, and R35GM148196, National Science Foundation grants DMS2052983 and IIS-1900473, Michigan State University Research Foundation, and Bristol-Myers Squibb 65109.
J.S. acknowledges funding from the NIH (R35GM148261).

\section*{Ethics declarations}
\subsection*{Competing interests}
The authors declare no competing interests.



\newpage

\bibliography{main_flag}

\end{document}